\documentclass[sigconf,nonacm]{acmart}
\setcopyright{none}

\usepackage{tabularx}
\usepackage{booktabs}
\usepackage{amsmath}
\usepackage{enumitem}
\usepackage{xcolor}
\usepackage{listings}
\hypersetup{hidelinks}

\usepackage{caption}
\usepackage{tikz}
\usepackage{pgfplots}
\pgfplotsset{compat=1.15}

\usetikzlibrary{positioning,shapes.geometric,arrows.meta,calc}

\lstdefinestyle{cuda}{
  language=C++,
  basicstyle=\ttfamily\scriptsize,
  breaklines=true,
  breakatwhitespace=false,
  postbreak=\mbox{\textcolor{gray}{$\hookrightarrow$}\space},
  columns=fullflexible,
  keepspaces=true,
  showstringspaces=false,
  tabsize=2,
  frame=single,
  framesep=3pt,
  xleftmargin=4pt,
  xrightmargin=4pt,
  commentstyle=\color{gray},
  keywordstyle=\color{blue!60!black},
}
\newcommand{\code}[1]{\texttt{#1}}

\AtBeginDocument{%
  }

\begin{document}

\title{From Custom-Fit to Portable: Bridging the Gap Between Synthesized and Engineered GPU Query Execution}

\author{Ivan Donchev Kabadzhov}
\affiliation{%
  \institution{EURECOM}
  \city{Biot}
  \country{France}
}
\email{ivan.kabadzhov@eurecom.fr}

\author{Eugenio Marinelli}
\affiliation{%
  \institution{EURECOM}
  \city{Biot}
  \country{France}
}
\email{eugenio.marinelli@eurecom.fr}

\author{Raja Appuswamy}
\affiliation{%
  \institution{EURECOM}
  \city{Biot}
  \country{France}
}
\email{raja.appuswamy@eurecom.fr}


\begin{abstract}
GPUs are increasingly used for analytical query processing, but developing GPU-based database engines that achieve the peak performance of the underlying hardware requires substantial research and engineering effort. A recent line of work argues that query processing should be synthesized, not engineered. In this scenario, instead of tuning a general-purpose engine to fit a workload, a large language model (LLM) generates code specialized to one query, one dataset, and one machine, thereby achieving an order-of-magnitude improvement in performance. This thesis, however, has so far been tested only on CPUs. In this work, we revisit the synthesize-versus-engineer debate for GPU analytics by answering three questions: (i) how good is synthesized GPU code?, (ii) why is it faster than engineered engines?, and (iii) how much of its advantage can be transferred back into a single, performance-portable engine?

To answer the first question, we present SHADB, an LLM-based synthesis framework that generates optimized CUDA or HIP kernels using an automated, profile-guided optimization loop. Using SHADB, we show that the synthesized code approaches the memory-bandwidth ceiling and outperforms a state-of-the-art JIT-compiled GPU database engine (HeavyDB) by 7.4$\times$ on SSB SF100. To answer the second question, we decompose this performance gap and systematically classify optimizations as generalizable or workload-specific. Finally, to answer the third question, we integrate these generalizable optimizations into SYCLDB, a performance-portable engine written entirely in the open SYCL programming model. Using optimized SYCLDB, we show that it is possible to substantially bridge the gap to synthesized code (within 1.27$\times$ total execution time) while retaining workload-level generality and hardware-level performance portability.
\end{abstract}

\maketitle

\subsection*{Artifact Availability Notice}
The implementation and evaluation details for our synthesis loop are maintained at our repository workspace (\url{https://gitlab.eurecom.fr/dislab/SHADB}).

\section{Introduction}
\label{sec:intro}

The past few years have witnessed a growth in adoption of GPUs for analytical workloads~\cite{Cao2023, crystal, GPUanalytics, GPUmicrosftDB}. However, the design of GPU databases is challenging due to the unique characteristics of GPU architectures, where fully exploiting the massive SIMT parallelism of GPUs requires attention to memory coalescing, occupancy, kernel-launch overhead, and inter-operator dataflow. Complicating this is the fact that GPUs from different vendors have different architectures, programming models, and performance characteristics, making it difficult to design a general-purpose GPU database engine that performs well across all workloads and hardware. 

A recent line of work argues that query processing should be \emph{synthesized, not engineered}~\cite{bespokeolap, gendb}. Instead of executing a query on a general-purpose engine whose operators are written once and tuned to be acceptable across every workload, these systems use large language models (LLMs) to generate code specialized to one query, one dataset, and one machine. For instance, GenDB synthesizes instance-optimized execution code through a chain of dedicated agents~\cite{gendb}. Bespoke OLAP~\cite{bespokeolap} generates workload-specific ``one-size-fits-one'' engines that outperform DuckDB~\cite{duckdb} by more than $10\times$~\cite{bespokeolap}. The common thesis is that the abstraction overhead a general engine pays is large enough that synthesizing query-specific code can provide substantial throughput/latency improvement especially under workloads with repetitive queries~\cite{saxena2024why}. 
However, all these systems target a \emph{CPU}. No prior work has investigated the synthesis--versus--engineering debate in the context of GPU databases. This work directly bridges the gap by answering the following questions: (1) How good is the synthesized code? (2) Why is it faster than engineered code? (3) Can optimizations used in synthesized code be generalized and applied to a general engine to close the performance gap?

Our central finding is that the ``synthesized--not--engineered'' thesis does not carry over from the CPU to the GPU for the star-schema workloads we consider in this work. On a CPU, a general engine's overhead is large enough that discarding generality for query-specific code can provide an order of magnitude improvement in performance. On a GPU, in contrast, this overhead is smaller, as engines are already optimized to exploit the massive parallelism of GPUs. Further, we find that most of the synthesized code's advantage comes from a set of optimizations that can be generalized and integrated into a portable engine. 
The residual performance edge that synthesized code has is small, non-portable (it does not survive even a change of dataset scale factor or GPU vendor), and expensive (${\sim}\$126$ and several GPU-hours to synthesize the benchmark). We therefore argue that, unlike on the CPU, engineering and not synthesis remains the right vehicle for high-performance GPU query processing, with LLMs serving instead as a tool for discovering the optimizations that a portable engine could adopt. In establishing this, we make the following contributions:

\begin{itemize}[leftmargin=*]
  \item We present SHADB (Synthesized Hardware Acceleration for GPU Databases), an LLM-driven synthesis framework that, given only a SQL query and its physical data layout, generates a self-contained CUDA (or HIP) program specialized to that query. Instead of relying on one-shot code generation, SHADB drives the LLM with an autonomous \emph{profile-guided optimization} (PGO) loop that audits each candidate for GPU-residency, compiles and validates it against a reference oracle, profiles it with hardware counters, and feeds the resulting bottleneck diagnosis back into the next prompt. This closed loop steers the synthesized code toward the memory-bandwidth roofline and lets SHADB establish an empirical performance ceiling. We evaluate SHADB on the Star Schema Benchmark (SSB)~\cite{ssb} and show that the synthesized code reaches the hardware ceiling and outperforms DuckDB~\cite{duckdb}, HeavyDB~\cite{heavydb},  and Crystal~\cite{crystal}  by $65\times$, $7.4\times$, and $2.6\times$ respectively.
  \item We perform an in-depth analysis to decompose the performance gap between the synthesized code and engineered code. In doing so, we make several interesting observations: (1) state-of-the-art kernel design techniques, like tiling, do not really provide a significant performance advantage over the simpler non-tiled approach adopted by LLMs, (2) the differences between synthesized and engineered code can be categorized into two types of optimizations, namely, those that can be generalized across queries and lifted back to a general-purpose engine (kernel fusion, query-specific coarsening, register-level inter-operator dataflow), and those that are specific to a workload (customized GPU primitives, lean group-by). 
  \item To determine whether we can bridge the gap between synthesized and engineered systems, this work builds upon SYCLDB~\cite{hal_2025}, a GPU-accelerated database engine written in the SYCL~\cite{sycl2020} programming model, with the goal of achieving performance portability across different GPU architectures. We start with an operator-at-a-time architecture which significantly lags behind the synthesized code and incrementally introduce the generalizable optimizations identified in the previous step. In doing so, we show how modern SYCL compilers enable a new implementation of kernel fusion (\emph{composition-based fusion}) which can be optimized to achieve register-resident, inter-operator dataflow without requiring the database engine to own a complex, compiler infrastructure (as required by all LLVM-based JIT compiling databases engines for lowering Intermediate Representation (IR) to device assembly.) 
  We show that these optimizations allow SYCLDB to reach within $1.27\times$ of the specialized code while answering every query from a single codebase and being portable across multivendor GPUs. 

\end{itemize}

\section{Synthesized versus Engineered GPU Acceleration}

In this section, we investigate the utility of LLMs in synthesizing GPU code. 
We first introduce SHADB. Then, we evaluate how well SHADB-synthesized GPU code performs against engineered analytical systems.

\subsection{SHADB: A Framework for LLM-Synthesized Database Kernels}
\label{sec:bg_synthesis}

\begin{figure}[t]
\centering
\usetikzlibrary{positioning,shapes.geometric,arrows.meta,calc}
\resizebox{0.85\columnwidth}{!}{%
\begin{tikzpicture}[
    node distance=0.8cm and 0.45cm,
    block/.style={draw, rectangle, rounded corners, minimum width=1.6cm, minimum height=0.7cm, align=center, fill=blue!5, draw=blue!40, font=\scriptsize},
    input/.style={draw, cylinder, shape border rotate=90, minimum width=1.0cm, minimum height=0.7cm, align=center, fill=gray!10, draw=gray!40, font=\scriptsize},
    gate/.style={draw, diamond, aspect=1.5, minimum width=1.3cm, align=center, fill=orange!5, draw=orange!40, font=\scriptsize},
    arrow/.style={-Latex, thick, draw=gray!70}
]
    \node[input] (input) {SQL Query\\\& Layout};
    \node[block, right=of input] (llm) {LLM Generative\\Engine};
    \node[block, right=of llm] (code) {C++/CUDA or\\HIP Source};
    \node[gate, right=of code] (audit) {Static\\Audits};
    \node[block, right=of audit] (nvcc) {NVCC/HIPCC\\Compiler};
    
    \node[block, below=1.0cm of nvcc] (gpu) {GPU Execution\\\& Correctness};
    \node[block, left=of gpu] (ncu) {Hardware\\Profiling};
    \node[block, left=of ncu] (diag) {Bottleneck\\Diagnosis};

    \path[arrow] (input) -- (llm);
    \path[arrow] (llm) -- (code);
    \path[arrow] (code) -- (audit);
    \path[arrow] (audit) -- node[above] {\tiny Pass} (nvcc);
    \path[arrow] (nvcc) -- (gpu);
    \path[arrow] (gpu) -- (ncu);
    \path[arrow] (ncu) -- (diag);
    
    \draw[arrow] (diag.west) -| node[pos=0.5, above] {\tiny Feedback} (llm.south);
    
    \draw[arrow, dashed] (audit.north) -- ++(0,0.4) -| node[pos=0.25, above] {\tiny Fail} (llm.north);

\end{tikzpicture}
}
\Description{A block diagram mapping SHADB's closed-loop pipeline where a SQL query and layout feed an LLM Generative Engine to synthesize source code that undergoes a Static Audit—looping back to the LLM if it fails—or passing to an NVCC/HIPCC Compiler, which executes on a GPU, runs through Hardware Profiling and Bottleneck Diagnosis, and returns performance feedback directly to the LLM to close the optimization loop.}
\caption{SHADB's closed-loop, PGO pipeline.}
\label{fig:shadb_architecture}
\end{figure}
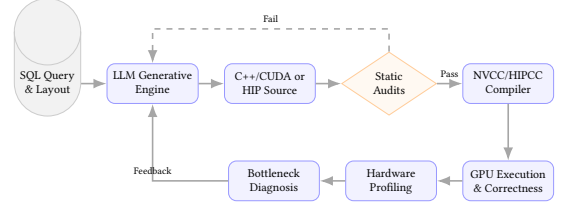


SHADB is an LLM-driven synthesis framework that, given a declarative SQL query and a physical dataset layout, synthesizes a self-contained C++ program with CUDA kernels for NVIDIA GPUs, or HIP kernels for AMD GPUs. Rather than lowering SQL into a fixed tree of reusable physical operators, it treats the whole query as the optimization unit, requiring the LLM to output a dual-payload containing a structured physical plan (such as column load, join, and aggregation layouts) and a matching executable source file. The physical plan is not an input to code generation, but plays an important role in SHADB for two reasons. First, committing to the physical strategy before emitting any C++ acts as a structured \emph{Chain-of-Thought} planning step~\cite{chain}, which reduces logical bugs and improves physical-design consistency relative to prompting for a raw kernel directly. Second, the plan is a declarative manifest of the model's intended design that can be audited. SHADB checks the emitted code against the declared strategy for contradictions, and uses the plan to compute an analytical roofline (the theoretical bytes moved per fact row) that it later compares against measured bandwidth to judge whether a kernel performs as intended. 
The generated code chooses data representations (such as direct-mapped dimension arrays and 1-byte packed L2-resident maps that we discuss later), join strategies, aggregation structures, kernel boundaries, and synchronization patterns specialized to the target query, scale factor, and device. We thus use SHADB to establish an empirical performance ceiling for query-specific GPU analytics against which we measure engineered and portable engines, and from which we later lift generalizable optimizations. 

Reaching the performance ceiling cannot be achieved with one-shot prompting, as LLMs cannot reliably emit near-roofline GPU code from the query text alone. SHADB therefore drives synthesis with an autonomous \emph{profile-guided optimization} (PGO) loop (Figure~\ref{fig:shadb_architecture}) that mirrors how a human performance engineer iterates (edit, compile, profile, and diagnose), with the LLM as the optimizer in the loop. Each iteration proceeds in stages: (1)~the model emits a dual payload consisting of a candidate plan and a matching self-contained C++/CUDA or HIP program; (2)~a static auditor enforces a strict GPU-residency contract, by checking the C++ code against the physical plan to catch contradictions (like the code using a host-side loop when the plan claims it runs entirely on the GPU) and rejecting broken versions
; (3)~survivors are compiled with the backend compiler (\texttt{nvcc} or \texttt{hipcc}) and (4)~verified against a reference oracle; (5)~verified candidates are profiled with hardware counters (\texttt{ncu} or \texttt{rocprof}) to measure DRAM throughput, L2 cache hit rate, occupancy, and register pressure; (6)~a bottleneck-diagnosis stage translates those counters into concrete, human-readable optimization directives (for example, raising memory-level parallelism or tiling a group-by into shared memory), which (7)~are injected as feedback into the next prompt. As a result, each optimized candidate is steered by measured hardware behavior, with the physical plan acting as the semantic bridge that closes the profile-guided optimization loop by matching that symptom  (like low L2 usage) against the declared strategy (for example, a direct-address lookup-table join) that tells the model precisely which architectural choice to revise in the next iteration. A candidate is adopted only if it both verifies and improves GPU execution time, and the loop continues until a fixed iteration budget is exhausted or no further gain is found. We call the first candidate to pass validation SHADB-Base and the converged, profile-tuned result SHADB-Opt. The gap between the two isolates what the optimization loop contributes.


\subsection{How Good is LLM-Synthesized Code?}
\label{sec:bg_how_good}

We compare SHADB with four analytical engines that span CPU execution, engineered GPU databases, portable GPU execution, and hand-specialized GPU kernels. DuckDB~\cite{duckdb} is a highly optimized vectorized CPU analytical database and serves as the engineered CPU baseline. HeavyDB~\cite{heavydb} is a mature GPU database that JIT-compiles query fragments through an LLVM-based backend targeting NVIDIA GPUs. SYCLDB~\cite{hal_2025}, in contrast, is a portable engine using an operator-at-a-time execution model written using the SYCL standard with the goal of supporting multiple vendors' GPUs. Crystal~\cite{crystal} is a  hand-written CUDA implementation of SQL queries and therefore approximates a manually specialized GPU lower bound without a query processing frontend. 
We evaluate SHADB in two forms to isolate the effect of its profile-guided optimization (PGO) loop, namely, SHADB-Base, the first synthesized candidate that passes validation, and SHADB-Opt, the result after the PGO loop has iterated to convergence.

All microbenchmarks, baseline systems, and end-to-end evaluations presented throughout this paper are executed on a unified heterogeneous server node (detailed in \S\ref{sec:eval-setup}) containing an Intel Xeon CPU, 1\,TB RAM, NVIDIA L40S, AMD MI210. We use the Star Schema Benchmark (SSB)~\cite{ssb} at scale factor 100 (SF100) as the primary workload, and report warm execution times averaged over 10 runs after a single warmup. To isolate pure processing performance, we report warm GPU kernel execution times for \textsc{SYCLDB} and HeavyDB (preloading data and using a warmup run to eliminate JIT compilation overhead). For DuckDB, we report the query processing time captured directly via its internal execution timer, which isolates the in-memory processing loop by excluding host-side parsing, optimization, and initial data I/O blocks. This establishes a fair baseline since neither \textsc{SHADB} nor Crystal contains a query management frontend. We only use SSB in this work because it is a well-established benchmark for analytical workloads that has been used extensively in other GPU database design literature~\cite{Yogatama2022, Yogatama2024}, and is the only benchmark supported by Crystal. SHADB was configured to use Anthropic's Claude Opus 4.8. Correctness of each synthesized CUDA program is checked against DuckDB as a reference oracle. DuckDB executes the query over the same raw columnar data and reduces the result to a single 64-bit integer (a plain integer sum for scalar queries, or a key-binding, per-group checksum modulo $2^{61}-1$ for grouped queries.) The synthesized CUDA program, performing the identical computation, is accepted only if its printed integer is bit-for-bit equal to the oracle's value.

Table~\ref{tab:llm-synthesized-code} reports warm SSB SF100 execution time and slowdown relative to SHADB-Opt. There are two important observations to be made. First, the LLM-synthesized code outperforms all analytical engines. 
Relative to SHADB-Opt, the engineered baselines are $7.8\times$ (SYCLDB), $7.4\times$ (HeavyDB), and $65.0\times$ (DuckDB) slower, respectively. SHADB-Opt even outperforms the hand-coded Crystal, providing $2.6\times$ improvement. This shows the potential of using synthesis for optimizing GPU query execution. Second, while the improvement from SHADB-Base to SHADB-Opt is modest ($69.81$\,ms to $60.47$\,ms over the eleven successfully-generated baselines), PGO plays two important roles: (i) it repairs failed candidates, such as Q3.1 and Q4.2 which are shown as the two \textsc{N/A} SHADB-Base entries, and (ii) it then extracts marginal gains once most optimized kernels are already near the memory roofline. For Q3.1, the baseline was rejected before compilation because the static source auditor detected a host-side aggregation loop (\texttt{host\_aggregate}), violating SHADB's GPU-only execution contract. A later iteration repaired the query as a fully GPU-fused kernel that matched the DuckDB oracle. For Q4.2, the baseline compiled but crashed during SF100 execution with a CUDA illegal-memory-access error in the generated kernel, indicating an out-of-bounds dimension lookup or aggregation access. The runtime error was fed back as repair context, and by the third PGO iteration the model generated a memory-safe implementation. 

\begin{table*}[t]
\centering
\caption{SSB SF100 execution time (warm, ms) and slowdown relative to SHADB-Opt of various systems.}
\label{tab:llm-synthesized-code}
\resizebox{0.8\textwidth}{!}{%
\begin{tabular}{lrrrrrr}
\toprule
Query & DuckDB & HeavyDB & SYCLDB & Crystal & SHADB-Base & SHADB-Opt \\
\midrule
Q1.1 & 404.0 (59.8$\times$) & 57.70 (8.5$\times$) & 58.36 (8.6$\times$) & 12.74 (1.9$\times$) &  7.66 (1.1$\times$) &  6.76 (1.0$\times$) \\
Q1.2 & 405.0 (106.0$\times$) & 53.00 (13.9$\times$) & 62.77 (16.4$\times$) & 12.73 (3.3$\times$) &  4.14 (1.1$\times$) &  3.82 (1.0$\times$) \\
Q1.3 & 385.0 (117.7$\times$) & 54.00 (16.5$\times$) & 61.32 (18.8$\times$) & 12.74 (3.9$\times$) &  3.79 (1.2$\times$) &  3.27 (1.0$\times$) \\
Q2.1 & 343.0 (57.7$\times$) & 39.60 (6.7$\times$) & 47.84 (8.1$\times$) & 13.30 (2.2$\times$) &  7.62 (1.3$\times$) &  5.94 (1.0$\times$) \\
Q2.2 & 354.0 (60.7$\times$) & 38.40 (6.6$\times$) & 46.88 (8.0$\times$) & 13.26 (2.3$\times$) &  8.45 (1.4$\times$) &  5.83 (1.0$\times$) \\
Q2.3 & 324.0 (70.9$\times$) & 38.40 (8.4$\times$) & 46.92 (10.3$\times$) & 13.16 (2.9$\times$) &  6.91 (1.5$\times$) &  4.57 (1.0$\times$) \\
Q3.1 & 495.0 (58.3$\times$) & 34.00 (4.0$\times$) & 53.79 (6.3$\times$) & 14.07 (1.7$\times$) & N/A &  8.49 (1.0$\times$) \\
Q3.2 & 365.0 (75.1$\times$) & 44.00 (9.1$\times$) & 21.31 (4.4$\times$) & 14.62 (3.0$\times$) &  5.24 (1.1$\times$) &  4.86 (1.0$\times$) \\
Q3.3 & 294.0 (74.6$\times$) & 43.60 (11.1$\times$) & 20.25 (5.1$\times$) & 13.78 (3.5$\times$) &  3.94 (1.0$\times$) &  3.94 (1.0$\times$) \\
Q3.4 & 252.0 (66.7$\times$) & 45.60 (12.1$\times$) & 20.21 (5.3$\times$) & 13.78 (3.6$\times$) &  3.78 (1.0$\times$) &  3.78 (1.0$\times$) \\
Q4.1 & 506.0 (49.4$\times$) & 46.00 (4.5$\times$) & 69.82 (6.8$\times$) & 22.23 (2.2$\times$) & 10.58 (1.0$\times$) & 10.24 (1.0$\times$) \\
Q4.2 & 496.0 (55.2$\times$) & 44.00 (4.9$\times$) & 60.45 (6.7$\times$) & 21.01 (2.3$\times$) & N/A &  8.98 (1.0$\times$) \\
Q4.3 & 446.0 (59.8$\times$) & 41.60 (5.6$\times$) & 34.14 (4.6$\times$) & 21.63 (2.9$\times$) &  7.70 (1.0$\times$) &  7.46 (1.0$\times$) \\
\midrule
Total & 5069.0 (65.0$\times$) & 579.90 (7.4$\times$) & 604.06 (7.8$\times$) & 199.05 (2.6$\times$) & 69.81$^\dagger$ (1.2$\times$) & 77.94 (1.0$\times$) \\
\bottomrule
\end{tabular}
}
\begin{minipage}{0.85\textwidth}
\footnotesize $^\dagger$SHADB-Base total and slowdown cover only the 11 queries with a valid baseline (Q3.1 and Q4.2 failed validation); the SHADB-Opt total over those same 11 queries is $60.47$\,ms.
\end{minipage}
\end{table*}


Profiling with \texttt{NCU} provides hardware-level explanation for the observed improvement. On NVIDIA L40S, 11 of the 13 queries achieve between $93.4\%$ and $97.0\%$ of the peak memory-bandwidth ceiling. 
These results establish SHADB as an empirical performance ceiling. Driven by its profile-guided loop, the LLM synthesizes query-specific GPU code that runs at the memory-bandwidth roofline and outperforms every engineered engine we test. That ceiling, however, is bought at a cost. First, generating the SHADB-Base across all queries takes 60.2\,minutes and \$10.06, while the full PGO loop takes 4.2\,hours and \$125.97. This is comparable to the costs reported by CPU-centric synthesis systems, with BespokeOLAP reporting 6--12 hours and \$120.99 for all TPC-H queries, and GenDB reporting 91 minutes and \$14.15 for five TPC-H queries. Second, each kernel produced by SHADB is tuned to a single query, scale factor, and device, and is not portable even across scale factors on the same benchmark. When we took code generated at SF10 and ran it unmodified at SF100, we produced correct answers in only two of thirteen queries. Five queries crashed on out-of-bounds accesses, and six silently returned wrong results. 

Portable engines such as SYCLDB sit at the opposite end of this trade-off, giving up peak performance for one reusable codebase that answers any query on any supported GPU. Whether these two extremes can be reconciled is the question the remainder of this paper investigates. We first decompose why the synthesized code is faster (\S\ref{sec:root}) in order to identify and separate generalizable optimizations that a portable engine can adopt from those fundamentally bound to a specific query or device. We then lift the generalizable ones into a portable engine (\S\ref{sec:bridge}) and measure how closely it can approach this ceiling (\S\ref{sec:eval}).

\section{Root cause analysis}
\label{sec:root}

In this section, we present an in-depth analysis of the performance gap between the synthesized CUDA kernels and the SYCLDB, HeavyDB, Crystal engines to answer the question: why is the synthesized kernel faster than these engineered systems? 

\subsection{Kernel Composition}
\label{subsec:kernel-composition}
The first difference across the systems we consider is how they compose kernels. A single SQL query is a directed acyclic graph of relational operators, and each engine has a different way of mapping that DAG to GPU kernels. SYCLDB is a ``kernel-at-a-time'' engine that runs one kernel per operator. For instance, a selection kernel writes a flag column, a projection kernel reads its inputs and writes a result column, a join is performed by build and probe kernels that build a hash table and probe it, etcetera. As a result, each intermediate output (selection flags and join results) is written to and re-read from GPU global memory. In contrast, the synthesized CUDA computes the entire query in a single straight-line pass, with intermediates held in registers and the base columns read once. This explains the large gap in performance between the synthesized kernels and SYCLDB.

Unlike SYCLDB, HeavyDB and Crystal do not employ a kernel-per-operator model. HeavyDB compiles a query by first generating a function as LLVM IR that runs once per input row. The body of the function is a chain of operations that decode the row's columns, apply predicates, compute projection arithmetic, and join/group-by key for the row. As these are all scalar operations that map one value to the next, HeavyDB fuses them into a single function. The function is then JIT compiled to GPU assembly (PTX). Fusion is, thus, automatic and code-generated, as the compiler emits operators as one straight-line function rather than as separate kernels. Despite this, HeavyDB is slower than the synthesized CUDA because it does not fuse the entire query into a single kernel. Operations that span shared, cross-row datastructures, are not generated and fused. Rather, they are implemented as general runtime routines. For instance, the join probe is implemented as a hash table lookup that reads the row's key, computes a hash, probes the hash table, and returns a pointer to the row's payload. The aggregation is implemented as a block-wide reduction that reads each thread's partial sum and atomically adds it to a shared/global sum. The SHADB generated kernel fuses the join probe and aggregation into the straight-line flow of the query, so that the join probe is a single load from a direct-mapped hash table and the aggregation is a single block-wide reduction. This explains why HeavyDB is slower than the synthesized CUDA.

Crystal, on the other hand, applies fusion manually as a library. Each query is a hand-written kernel that fuses the whole query into one specialized, register-resident kernel, including the join probe and the aggregation. The data structures are part of the fused register flow. Thus, from the fusion point of view, Crystal is more similar to the synthesized CUDA than HeavyDB. Despite this, Crystal is slower than the synthesized CUDA. The rest of this section breaks down the difference by studying both workload-agnostic (execution model) and workload-specific (optimized primitives and dataflow) factors that contribute to the performance gap.

\subsection{Execution Model}

A key difference between Crystal and the synthesized CUDA is with respect to the execution model. 

\paragraph{Crystal: the tile model.} A key characteristic of Crystal is its \emph{tile}-based execution model. This model is based on the observation that threads within a thread block in CUDA can communicate through shared memory and can synchronize through barriers. A single thread block can hold a significantly larger group of elements collectively between them in shared memory compared to a single thread. This new granularity of execution unit is referred to as a \emph{tile}. In the tile-based execution model, instead of viewing each thread as an independent execution unit, we view a thread block as the basic execution unit with each work group explicitly processing a tile of entries at a time. Crystal builds a library of block-wide primitives that operate on a tile of rows, including \code{BlockLoad}, block scan and compaction, \code{BlockReduce}, and hash-table probes. Each kernel is written to operate on a fixed-size tile of rows, and the grid is sized to have one block per tile. Each block loads the tile into registers and runs block-wide primitives over it. This design was supposed to confer three benefits. First, staging a tile of rows into per-thread registers (and shared memory) lets every operator in the query act on the \emph{same} in-register tile, so intermediates are fused on-chip and never round-trip through global memory. Second, the primitives load a tile so that consecutive threads touch consecutive elements, yielding fully coalesced 128-byte transactions and thus near-peak effective bandwidth. Third, because each block processes a bounded, fixed-size tile, the per-block working set fits on-chip and the block count scales with the data, leaving load balancing to the GPU scheduler. Listing~\ref{lst:crystal-tile} shows an example for Crystal's SSB Q11 scan. As can be seen, the grid is \emph{one block per fixed-size tile} (\code{<128,4>} ${}={}$ a 512-row tile, hence \code{ceil(n/512)} blocks), and each block loads a register tile of consecutive items and runs block-wide primitives over it.

\begin{lstlisting}[
  float=t,
  caption={Crystal tile-model kernel (one block per 512-row tile).
           \code{ull} denotes \code{unsigned long long}.},
  label={lst:crystal-tile},
  language=C++,
  basicstyle=\ttfamily\scriptsize,
  numbers=none,
]
// ONE block per 512-row tile (<128,4>).
// BT=block threads, IPT=items per thread; 
template<int BT,int IPT>           // <128,4>=512 rows
__global__ void scan(int*date,int*disc,int*qty,
                     int*ext,int n,ull*revenue){
  int items[IPT], items2[IPT], flags[IPT]; // REGISTER tile
  int off = blockIdx.x*TILE_SIZE;  // this block's tile
  BlockLoad(date+off,items,n);     // load WHOLE tile,
  BlockPredGT   (items,19930000,flags,n); // then test
  BlockPredAndLT(items,19940000,flags,n); // every item:
  BlockLoad(qty +off,items,n);            // no early-out
  BlockPredAndLT (items,25,flags,n);
  BlockLoad(disc+off,items,n);
  BlockPredAndGTE(items, 1,flags,n);
  BlockLoad(ext +off,items2,n);    // payload ALWAYS loaded
  long long sum=0;
  for(int k=0;k<IPT;++k)
    if(flags[k]) sum+=items[k]*items2[k];   // masked
  if(!threadIdx.x) atomicAdd(revenue,BlockSum(sum));
}
// grid=ceil(n/TILE_SIZE): one block per tile; block=128
\end{lstlisting}


\paragraph{SHADB: grid-stride persistent threads.} SHADB, in contrast, launches a fixed, machine-sized grid with enough blocks to fill the GPU hardware units. Each thread loops over a large sequence of elements, advancing its pointer by the total grid thread count at each iteration. Listing~\ref{lst:shadb-q11} illustrates this using the SSB Q11 scan implementation. The grid is dynamically sized to fully fill the execution pipeline (\code{maxActiveBlocksPerMultiprocessor} $\times$ \code{numSM}, capped at a safe machine constraint such as \code{GRID\_CAP}${}={}262144$). 

The persistent threads loop across the target table bounds, allowing thread $t$ to handle rows $t$, $t \mathbin{\text{+}} S,\ t \mathbin{\text{+}} 2S,\dots$, where the stride $S$ matches the exact global thread space (\code{blockDim.x * gridDim.x}). 

This execution pattern yields three critical advantages. First, within any single execution loop iteration, consecutive threads access consecutive array positions. This ensures that the warp-level operations perfectly coalesce into wide 128-byte hardware transactions, preserving near-peak DRAM efficiency. The strided index jumps are isolated \emph{between} loop boundaries, eliminating divergence within the active warp. Second, by structuring the table traversal inside a continuous grid-stride loop, the individual steps remain completely independent. This exposes opportunities for the backend compiler to unroll the execution blocks, pipelining vector memory reads (such as loading wide \code{int4} structures) to maximize memory-level parallelism (MLP) and fully saturate the GPU pipeline. Third, utilizing persistent threads allows each processing unit to safely maintain a local register partial accumulator (\code{local}) across its entire assigned row sequence, executing only a single global atomic addition (\code{atomicAdd}) upon loop termination rather than contending on global variables for every matching tuple.

\begin{lstlisting}[
  float=t,
  caption={SHADB grid-stride persistent-thread scan (q11).
           \code{ull} denotes \code{unsigned long long}.},
  label={lst:shadb-q11},
  language=C++,
  basicstyle=\ttfamily\scriptsize,
  numbers=none,
]
// grid FILLS the GPU; each thread strides over rows.
__global__ void q11(const int4*od4,const int*disc,
     const int*qty,const int*ext,long long n,ull*result){
  ull local=0;                     // REGISTER accumulator
  const long long stride =         // = total #threads
        (long long)blockDim.x*gridDim.x;
  for(long long g=blockIdx.x*blockDim.x+threadIdx.x;
      4*g<n; g+=stride){           // jump by whole grid
    int4 v=od4[g];                 // ONE 128b load = 4 rows
    int od[4]={v.x,v.y,v.z,v.w};
    for(int l=0;l<4;++l){
      long long i=4*g+l; if(i>=n) break;
      if(od[l]<OD_LO||od[l]>OD_HI) continue; // miss: skip
      int d=disc[i];               // read iff passed
      if(d>=1&&d<=3&&qty[i]<25)
        local+=(long long)ext[i]*d;
    }
  }
  // block-reduce local, then ONE atomic per block:
  if(!threadIdx.x) atomicAdd(result,blocksum);
}
// block=256; grid sized to fill all SMs (persistent)
\end{lstlisting}



Given the above, the first question that arises is whether the tile abstraction itself can lead to worse performance and explain Crystal's performance gap with the synthesized CUDA. We test this hypothesis by modifying Crystal and implementing a \emph{direct} kernel that does not use tiling or any of the block primitives. Table~\ref{tab:tiling-ab} reports the result when both kernels are run on an NVIDIA~L40S (Ada, \texttt{sm\_89}, ${\approx}864$~GB/s DRAM peak, $48$~MB L2).

\begin{table}[t]
\centering
\caption{Microbenchmark isolating the tile abstraction (NVIDIA~L40S,
warm-min kernel ms; DRAM as \% of peak; L2-hit\% for the join probe).}
\label{tab:tiling-ab}
\small
\resizebox{0.85\columnwidth}{!}{%
\begin{tabular}{llrrr}
\toprule
Query & Variant & Time (ms) & DRAM\% & L2\% \\
\midrule
Q1.1 (scan) & block (Crystal tile)      & 12.74 & 97.2 & --- \\
            & direct (no tiling)          & 13.01 & 94.9 & --- \\
\midrule
Q3.1 (join) & block (tile)              & 14.15 & 65.3 & 70.1 \\
            & direct (no tiling) & 12.37 & 44.6 & 80.3 \\
\bottomrule
\end{tabular}%
}
\end{table}

\paragraph{Bandwidth-bound scan (Q1.1).} The block tile and the direct kernels complete the query in $12.74$ vs $13.01$~ms at SF100. Both drive the L40S's ${\approx}864$~GB/s DRAM at $95\text{--}97\%$ of peak. The direct kernel in fact does roughly twice the compute (SM throughput $36\%$ vs $15\%$) and still ties, because memory bandwidth is the bottleneck and the compute is hidden behind it. The block primitives do not provide any benefit on the bandwidth-bound scan.

\paragraph{Latency-hidden join (Q3.1).} For join queries, the tile abstraction and block primitives surprisingly lead to a deterioration in performance. Table~\ref{tab:tiling-ab} shows the result for Q3.1 where the direct kernel ($12.37$~ms) outperforms the tiling approach ($14.15$~ms) at SF100. 
The reason behind this is the fact that Crystal's block primitives \texttt{BlockLoad} a full tile of every referenced column unconditionally and only then mask with \texttt{selection\_flags}. So Crystal reads every referenced column for every row regardless of selectivity. The synthesized kernels short-circuit this process, evaluating predicates first and never issuing the loads of the later columns for rows already eliminated. 

The Q1.x scans make this directly measurable. Both kernels read the same four \code{lineorder} columns, so Crystal moves a fixed $600{,}043{,}265 \times 4 \times 4\,\text{B} = 9.60$\,GB on every one of them. Profiled with \code{ncu}, the synthesized kernel instead moves only $5.71$, $3.23$, and $3.00$\,GB on Q1.1, Q1.2, and Q1.3, which is $1.68\times$,
$2.97\times$, and $3.20\times$ less. Data moves shrink monotonically as the query grows more selective and the kernel skips the aggregate-column loads for filtered rows. Both kernels are DRAM-bandwidth-bound---each sustains $87\text{--}97\%$ of the L40S's ${\approx}864$\,GB/s peak. So this difference in data read translates directly into a runtime difference, as Crystal is $1.9\text{--}3.9\times$ slower across the three scans, closely tracking the $1.7\text{--}3.2\times$ gap in bytes moved. The same mechanism operates on the join flights. Crystal is only $1.66\times$ slower on Q3.1, because with little filtering both tiled and synthesized kernels read essentially everything, while the more selective Q3.3 and Q3.4 reopen the gap ($3.5\times$ and $3.6\times$). 

This microbenchmark reveals a previously unidentified negative result: tiling does not provide any benefit under scans that are bandwidth-bound and is a net penalty in joins. A direct, non-tiled kernel matches or beats it in all cases in our setup as it results in substantially fewer bytes being moved.

\subsection{Workload-Specific Optimizations}
\label{subsec:workload-opt}
Having discussed the workload-agnostic execution model, we now turn to the workload-specific optimizations that SHADB applies to the SSB queries. The Star Schema Benchmark groups its 13 queries into four \emph{flights}, each a distinct execution shape. 
SHADB applies the following optimizations that are shape specific, none of which are present in Crystal.

\paragraph{Direct-mapped dimension tables and packed probe${}+{}$gather.} Crystal builds a modulo-hashed,
double-sized open hash table with collision handling that is populated by multiple threads with \code{atomicCAS}. Probing is done by recomputing the hash and testing for empty slots. SHADB exploits SSB's dense primary keys by building a directly-indexed \code{map}. As a result, the build is a race-free direct store with no atomics or CAS, and the probe is just one data load with no modulo and no collision handling for each of the 600\,M fact rows in SF100. This boosts all queries that involve a join (\emph{Q2.x}, \emph{Q3.x}, \emph{Q4.x}). As queries \emph{Q1.x} do not have any joins, it does not impact them.

\paragraph{Byte-packed, L2-resident dimension maps.}
Queries in flight Q4.x join four dimension tables. Consider Q4.1:
\begin{lstlisting}
SELECT d_year, c_nation, SUM(lo_revenue - lo_supplycost) AS profit
FROM   lineorder, customer, supplier, part, ddate
WHERE  lo_custkey = c_custkey AND lo_suppkey = s_suppkey
  AND  lo_partkey = p_partkey AND lo_orderdate = d_datekey
  AND  c_region = 'AMERICA' AND s_region = 'AMERICA'
  AND  (p_mfgr = 'MFGR#1' OR p_mfgr = 'MFGR#2')
GROUP BY d_year, c_nation;
\end{lstlisting}
Here the 3M-row \code{customer} dimension table is both \emph{filtered} (on \code{c\_region})
and \emph{gathered} (\code{c\_nation} in the GROUP BY). A direct-mapped table
indexed by \code{c\_custkey} that stores the nation as a 4-byte \code{int} needs
$3\text{M}\times 4 = 12\,\text{MB}$, in addition to a separate region-filter array. SHADB
fuses both into a single \emph{1-byte} map: \code{nat[c\_custkey]} holds the nation
code, which lies in $\text{[}0,24\text{]}$ and fits in a byte when \code{c\_region} matches, and
the reserved sentinel \code{0xFF} otherwise. A probe is then one byte load that at
once tests the filter and returns the group value. As a result, the customer working set
shrinks from 12MB to 3MB and the customer probe
over the ${\sim}120$\,M surviving rows stays resident in the L40S's $48$\,MB L2 cache even
while the supplier, part, and date maps compete for the same cache. Q4.2 and Q4.3,
which filter but do not do a GROUP BY, likewise pack it into a 1-byte flag.
Crystal, in contrast, keeps \code{customer} in its $24\,\text{MB}$ modulo-hashed table and cannot
pack it at all. 

\paragraph{Shared-memory privatized group-by.}
When a query has a grouped \code{SUM}, Crystal scatters every surviving fact row into a global group table with an
\code{atomicAdd}. When the table is \emph{small} and many rows survive, those atomics
all contend on a few counters. Q4.1 (above) is the extreme case: its
\code{GROUP BY d\_year, c\_nation} has only 175 groups, yet
${\sim}120$\,M rows survive to update them. SHADB privatizes the table by letting each thread block
use its own 175-entry copy in shared memory, accumulates into it with
shared-memory atomics during the scan, and issues just one global \code{atomicAdd}
per group when the block finishes. This cuts global atomic traffic from
${\sim}120$\,M updates to at most $175$ per block. Q2.2
(\code{GROUP BY d\_year, p\_brand1}, $56$ groups) uses the same scheme. This optimization pays only when the group count fits shared memory and contention is high enough. So SHADB does not apply it to other queries, like Q3.2 and Q4.3 which group by \code{city}$\times$\code{city} /
\code{city}$\times$\code{brand}. In this case, there are hundreds of thousands of groups that do not fit in shared memory. On these queries, SHADB keeps the global table because too few rows survive for contention to matter. Similarly, on  selective join queries where very few fact rows are grouped, this optimization has no effect. The Q4.2 kernel, for
instance, resorts to a global table as only ${\sim}0.46\%$ rows survive, leading to low contention.

\section{Bridging the gap}
\label{sec:bridge}

In this section, we answer the third question raised in \S\ref{sec:intro}: how close can a performance-portable engine get to the SHADB roof? In this work, we build on SYCLDB, a performance-portable engine written entirely in the cross-architecture, open-standard programming model SYCL. As explained earlier, SYCLDB represents the opposite end of the generality spectrum; while SHADB customizes code to a specific query and hardware, SYCLDB~\cite{hal_2025} aims to be fully portable across workloads and multi-vendor hardware. This section first provides an overview of the baseline, operator-at-a-time SYCLDB. Then, we explore which workload optimizations used by SHADB can be generalized, and describe how we lift and apply them to SYCLDB.

\subsection{Baseline SYCLDB}
\label{subsec:baseline-sycldb}

The SYCLDB we use as our starting point expresses each relational operator as a data-parallel SYCL kernel. Each kernel is written as a \code{parallel\_for} over a one-dimensional range with one work-item per fact row, leaving the mapping of work-items onto work-groups and compute units to the SYCL runtime. Thus, the execution model in SYCLDB is the opposite of Crystal's \emph{tile} model (Listing~\ref{lst:crystal-tile}), where the kernel is templated on a fixed tile shape and each work-group stages a register tile of consecutive rows. SYCLDB writes no tile shape and sizes no grid, so the identical kernel lowers to any vendor's GPU with the work-item-to-hardware mapping left to the compiler. The listing below shows an example selection over \code{n} fact rows.
\begin{lstlisting}
q.parallel_for(sycl::range<1>(n), [=](sycl::id<1> i){
    // global memory read and write
    flags[i] = flags[i] && (col[i] >= lo && col[i] <= hi);
});
\end{lstlisting}

Without fusion, this model runs one such kernel per operator. The selection above streams a fact column together with the \code{flags} array, ANDs its predicate into \code{flags}, and writes the array back. The next predicate
reads \code{flags}, combines its own result, and writes it again. 
A query whose fact-side query plan has $N$ such operators will therefore launch $N$ kernels and make $N$ streaming passes over the fact arrays, materializing and re-reading the intermediate flags array between every pass. Because the engine is bandwidth-bound, memory access is a major source of overhead for SYCLDB. As HeavyDB and SHADB both adopt fusion, albeit in different ways (code-generated versus synthesized), collapsing the chain into a single pass with fusion is the first optimization we added to SYCLDB.

\subsection{Fusion}

\subsubsection{The dynamic-function mechanism.}

The canonical \emph{code-generation} approach for fusing GPU kernels adopted by systems like HeavyDB derives directly from the CPU code generation approach proposed by Neumann et al. in Hyper~\cite{Neumann2011}. With this approach, at query time, a code generator walks the plan and emits fresh IR, expressing each per-row operator as values in one specialized per-query function. Standard compiler passes (inlining, scalar-replacement, etcetera) promote this function to register-resident code, which is then lowered to the target instruction set architecture (ISA), such as PTX for NVIDIA GPUs. Thus, a new, fused function body is synthesized for every query shape with code-generation-based fusion.

Recently, SYCL compilers have introduced new functionality that enables applications to provide fusion hints at a higher level without having to explicitly handle various code-generation aspects. AdaptiveCpp's \textbf{dynamic function} mechanism is one such example that enables fusion of a list of \emph{operators}. We provide a brief overview here to introduce terminology and follow up with a detailed example below. Each fusable operator is written once as an ordinary SYCL device function and compiled ahead-of-time (when the application binaries are built) as portable IR. All fused operators share a common signature and are referred to as \emph{slots}. The application also declares a \emph{hole} function at compile time. At runtime, the application can decide which slots to use dynamically to fill the hole function. AdaptiveCpp JIT \emph{composes} a fused kernel by using the chosen slot symbols to fill the pre-declared \emph{hole} function as a call sequence and lowering the result to device assembly.
Thus, while a code-generation-based approach emits completely new IR per query, this \emph{composition-based fusion} approach splices a runtime-chosen sequence of pre-compiled IR symbols. 

A direct consequence is that a code-generation engine must build and maintain a full IR emitter  whereas a SYCL composition engine writes its operators as ordinary SYCL/C++ device functions and lets the SYCL compiler generate the IR from that C++, and the runtime lowers it to the target ISA. The engine therefore never emits IR itself. This, in turn, has the effect that a composition-based fusion approach is more portable than a code-generation-based one. HeavyDB's generator emits IR for LLVM's NVPTX backend, so its fused kernels are CUDA specific, whereas a SYCL composition engine compiles its slots to portable IR and thus inherits, for free, portability across any vendor's GPU supported by the compiler/runtime. A code generator could match this only by additionally targeting a portable IR such as SPIR-V, while composition gets it by construction.

\subsubsection{Fusion in SYCLDB}
Given the mechanism above, generating a fused kernel in SYCLDB reduces to defining the slot function ABI and determining which slot sequences should fill the \code{hole} function for any given physical plan. We illustrate the process of how fusion works in SYCLDB end-to-end with a representative query given below.

\begin{lstlisting}[language=SQL]
SELECT d_year, SUM(lo_revenue)
FROM   lineorder, ddate
WHERE  lo_orderdate = d_datekey      -- join
  AND  lo_discount BETWEEN 1 AND 3   -- fact filter
GROUP BY d_year;
\end{lstlisting}
\noindent for which Calcite emits the physical plan below; its columns occupy a single
\emph{global ordinal} space formed by concatenating the scanned tables
(\code{lineorder}~$=0\text{--}16$, \code{ddate}~$=17\text{--}33$):
\begin{lstlisting}
#0 TABLE_SCAN lineorder
#1 TABLE_SCAN ddate
#2 JOIN inner on =($5,$17)        -- lo_orderdate = d_datekey
#3 FILTER     $6>=1 AND $6<=3     -- lo_discount BETWEEN 1 AND 3
#4 AGGREGATE  group=[21] SUM($12) -- d_year (17+4), SUM(lo_revenue)
\end{lstlisting}

The fused kernel is built from the four pieces shown in Listing~\ref{lst:sycl-fusion}, corresponding to the example query. The slot \emph{library} (part~1) supplies the operator types. Each slot is one fusable operator, written once as an ordinary device function and compiled ahead-of-time into the device image. So that slots can be chained
uniformly, every slot shares one signature. In the basic ABI this is simply
\code{(size\_t i, JitCtx c)}, where \code{i} is the fact-row index and \code{JitCtx} is
a small struct of data containing column pointers, literals, and op-codes, through which
the executor tells each slot which columns and constants to act on. The query's entire
identity lives in this \code{JitCtx} data, not in any generated code.

Because the \code{JitCtx} is the only thing slots share, it is also the channel by
which one operator hands a value to a later one. The executor therefore places two
scratch arrays in global memory and exposes them through the \code{JitCtx}, namely, a per-row
selection flag, \code{pass}, and a set of result buffers \code{res} for gathered
values. A selection writes \code{c.pass[i]}; a join probe clears \code{c.pass[i]} on a
miss; a gather deposits its dimension value into \code{c.res}; and the terminal reads
both back. In this basic ABI, every hand-off between operators is thus a write and a
later read of a global array.

\begin{lstlisting}[caption={SYCLDB kernel fusion via dynamic functions}, label={lst:sycl-fusion}]
// (1) Every operator is a generic device slot, one shared signature.
SYCL_EXTERNAL void jit_sel_lit_0(size_t i, JitCtx c);
SYCL_EXTERNAL void jit_probe_2  (size_t i, JitCtx c);
SYCL_EXTERNAL void jit_gather_3 (size_t i, JitCtx c);
// ... fixed slot kinds, each replicated for up to 16 chain positions

// (2) The HOLE: only declared; the JIT fills it with a slot sequence.
SYCL_EXTERNAL void jit_dispatch(size_t i, JitCtx c);

// (3) One fused kernel: run the chain, then the terminal.
q.parallel_for(range, [=](size_t i){
    ctx.pass[i] = true;
    jit_dispatch(i, ctx);
    if (ctx.pass[i]) terminal(i, ctx);
});

// (4) The executor composes the chain, then lowers it.
acpp_jit::dynamic_function_config cfg;
cfg.define_as_call_sequence(&jit_dispatch,
      { jit_sel_lit_slot(0), jit_sel_lit_slot(1),
        jit_probe_slot(2), jit_gather_slot(3) });
auto kernel = cfg.apply(/* wrapper in (3) */);
q.parallel_for(range, kernel);
\end{lstlisting}

Parts 2 and 3 are also fixed. The \emph{hole} function in part~2 is \code{jit\_dispatch} in Listing~\ref{lst:sycl-fusion}. It is a function that is declared but never defined that the kernel body calls as a stand-in for the fused operator chain. \emph{Dispatch} is the act of filling that hole. At submit time the
executor binds \code{jit\_dispatch} to a concrete \emph{call sequence} of slots via
\code{define\_as\_call\_sequence}, and the JIT inlines them. Finally, the
\emph{terminal}, which is the aggregate or group-by, is not a slot. It lives in the kernel
\emph{wrapper} (part~3) and consumes \code{c.pass[i]} once the chain has run. In the example, the
wrapper runs row \code{i} through \code{jit\_dispatch} and, if \code{c.pass[i]} is still
set, calls the terminal that updates the \code{d\_year} group. 

Only part~4 is query-specific as it contains the call sequence the executor composes for this plan. To fill part~4, the executor first has to identify the operators that need to be added to slots. SYCLDB does this by keeping the fact table as the spine of execution and treating every fusable operator as a transformation over fact rows; a selection
narrows \code{pass}, a join probe drops non-matching rows, a gather pulls one
dimension attribute into the row, and the terminal consumes the survivors. Walking the
plan, the executor identifies operators applied to the fact table (\code{lineorder} in the running example) and lowers each operator of the chain to one slot,
recording its operands (column pointers, literals, op-codes) in the corresponding
\code{JitCtx} position. For the example it emits the following slots: (i) two \code{jit\_sel\_lit} slots for the conjunctive predicate (the \code{BETWEEN} becomes the AND of its two bounds, each narrowing \code{pass}), (ii) a \code{jit\_probe} slot that looks up
\code{lo\_orderdate} in the \code{ddate} table, setting \code{pass} false on a miss and
recording the matched dimension row on a hit, and (iii) a \code{jit\_gather} slot that
materializes \code{d\_year} for the matched row. Note that the dimension setup that involves building the probe table for dimension columns (like \code{ddate} in the example) is done by a separate kernel and not fused.


With the sequence chosen, \code{define\_as\_call\_sequence} shown in part~4 binds it into
\code{jit\_dispatch} and \code{cfg.apply} lowers the wrapper for the target device. After JIT, this one call to \code{jit\_dispatch} inlines the entire operator chain including the selections, the join probe(s), and the gather(s). In doing so, the inlined code (a) clears \code{pass} if the row fails a predicate or misses a probe, and (b) stashes any gathered dimension values in \code{ctx}. The terminal then computes the \code{GROUP BY} for \code{d\_year} for rows that survive the whole chain. Thus, the chain is built by hole composition (runtime-chosen slots), while the terminal is hand-written in the wrapper. Both end up inlined into one kernel that makes one pass over the fact rows. So the whole query (chain + terminal) is fused. 


 \subsubsection{Register table optimization.}

 The basic ABI fuses the operators' control flow into one kernel but not their data flow. As slots share only the global \code{JitCtx}, every access to the gathered value is stored to the \code{JitCtx.res} buffer by its producer and re-loaded by the terminal. This results in a full DRAM round-trip even though both are inlined into the same kernel. A code-generation-based fusion approach would store these values directly in registers and avoid the DRAM access. Thus, in practice, SYCLDB uses a different slot that passes the \code{pass} flag and gather values by reference---\code{(size\_t i, JitCtx ctx, RegFile \&rf)}, where \code{RegFile} is a tiny per-thread array (\code{rf.v}) containing the gathered values. As long as accesses to \code{rf} are compile-time constants, the compiler will store the values in a register. This implies that both the gather's write to \code{rf.v} and the terminal's read from it must use compile-time constant indexes. 
 
 We ensure this constant index at both the gather that writes \code{rf.v}, and the terminal that reads it, with two complementary mechanisms. On the write side, a macro expands each operator body into one symbol per chain position, encoding the slot number as a literal. Listing~\ref{lst:sycl-fusion} composes the gather in plan order as \code{jit\_gather\_3} (slot~3), whose body therefore writes \code{rf.v[3]} with a constant index, the executor selecting this symbol because the gather lands in slot~3. On the read side, the terminal must read each group key back from \code{rf.v[j]} but cannot know which slot produced it. So the executor \emph{reorders} the chain to place the \code{K} gathered group keys in the first \code{K} slots, in group order. The terminal then reads key~0 from \code{rf.v[0]}, key~1 from \code{rf.v[1]}, and so on, creating a fixed, compile-time pattern independent of the query. In the example query, there is a single group key, \code{d\_year}, so its gather is renumbered to slot~0 (\code{jit\_gather\_0}, writing \code{rf.v[0]}) and the terminal reads it at the constant index~0. The reorder is a pure renumbering of the same operators and leaves the result unchanged. We refer to this optimization as the \emph{register table} optimization.

\subsection{Lifting Workload-Specific Optimizations into the Engine}

\noindent The root-cause analysis (\S\ref{sec:root}) attributes the synthesized
code's advantage to a stack of optimizations.
We now revisit them to identify generalizations of these optimizations that can be lifted into SYCLDB while retaining performance portability, and workload-specific optimizations cannot be lifted. 


\subsubsection{Optimizations lifted into the engine}


\noindent\textbf{One-load packed probe${}+{}$gather.}
As mentioned in \S\ref{subsec:workload-opt}, the synthesized code uses direct-mapped dimension tables and packed probes so a star join that needs that column costs a single load. This optimization relies on the assumption that primary keys are dense enough to index directly, which is the case in SSB. We lift this optimization into SYCLDB directly by implementing new build and probe primitives that build a packed table and probe a packed slot. At runtime, SYCLDB verifies the key range and chooses either the packed version, or falls back to the default unpacked version. As we show later, this optimization helps low-selectivity join-heavy queries with many probe matches and has little effect on the very selective queries where few rows survive to be gathered.

\noindent\textbf{Asymmetric batching.}
A second optimization used by both SHADB and Crystal is to have each thread process
several fact rows per iteration, raising memory-level parallelism (MLP). A general engine can neither hard-code one items-per-thread value nor afford to search
for it per query. Our analysis (\S\ref{sec:root}) showed that streaming scan issues several independent column loads that batching keeps in flight at once, whereas a join creates many dependent dimension lookups. We use this insight in SYCLDB to apply batching \emph{asymmetrically} by choosing the batch size based on the terminal. Recall that the terminal is the fused kernel's final operation (aggregate or \code{GROUP BY}). There are two terminal types, namely, the \emph{scalar scan--reduce} terminal (for example, a scan of
\code{lineorder}, a few predicates, and a single \code{SUM}) and the \emph{join/group} terminal (for example, a scan of
\code{lineorder}, probes into the dimension tables, a gather of the group keys, and a
grouped \code{SUM}). SYCLDB uses a batch size of $4$ if the query uses a scan terminal, and $1$ if there is a join terminal. Crucially, this factor governs the \emph{entire} fused loop, not just the terminal. Each thread runs the whole operator chain once per batched row, so the fused operators
also receive the benefit of batching. Because the terminal type already encodes the query shape, every query inherits the right batch size with no per-query tuning.

\subsubsection{Optimizations left out}

\noindent Two of SHADB's optimizations are \emph{workload-} or
\emph{hardware-specific} and do not generalize. The \textbf{byte-packed,
L2-resident dimension map} fuses one particular dimension's filter and group column
into a single one-byte map to keep that table resident in L2. This makes it tied to specific
columns and cardinalities of one schema. The \textbf{shared-memory privatized
group-by} accumulates into a block-local table and is valid only when the group
count fits shared memory, which holds for some queries and not others. SYCLDB, in contrast, keeps
a global table. 

\section{Evaluation}
\label{sec:eval}

In this section, we present an evaluation of the optimized SYCLDB to answer the following questions: (i) how does SYCLDB perform in comparison to SHADB and other CPU/GPU engines? (ii) how portable is the performance of SYCLDB to heterogeneous hardware?, (iii) what is the contribution of each individual optimization in SYCLDB? 

\subsection{Experimental Setup}
\label{sec:eval-setup}

\textbf{Hardware Platforms.} We evaluate all systems on a unified heterogeneous server that contains dual-socket Intel Xeon Platinum 8454H processors (Sapphire Rapids, 64 cores, 128 threads total, 3.4GHz boost), 1\,TB DDR5 ECC host memory, and a Gen4 NVMe SSD (7,400\,MB/s sequential reads). The accelerator array consists of two distinct GPU architectures connected via PCIe Gen4 x16 interconnects: (i) NVIDIA L40S: Ada Lovelace architecture, 48\,GB GDDR6 VRAM, 864\,GB/s peak bandwidth; (ii) AMD Instinct MI210: CDNA 2 architecture, 64\,GB HBM2e VRAM, 1.6\,TB/s peak bandwidth.

\textbf{Software Stack.} We use AdaptiveCpp 25.10.0 (compiled via LLVM 18) as our SYCL toolchain, CUDA 12.6 SDK backend for the NVIDIA GPU, and the ROCm 6.3.3 HIP backend for the AMD GPU. For baseline comparisons, DuckDB runs entirely on the host CPU. Hand-written Crystal kernels are compiled natively using NVCC. 
HeavyDB is executed inside a containerized environment pinned to the ~\texttt{omnisci/core-os-cuda:latest} (5.10.2) image, which was pulled directly from the official Docker Hub repository since the project's original GitHub links are no longer accessible.
SHADB was configured to use Anthropic's Claude Opus 4.8. We described the SHADB configuration in detail in \S\ref{sec:bg_how_good}.

\textbf{Workload \& Protocol.} Experiments are conducted using the SSB benchmark at SF100 ($\sim$600M fact rows) and SF200 ($\sim$1.2B fact rows). For each system, we preload the input columns into the database engine  to isolate device execution from storage/PCIe transfer latency. To eliminate transient JIT or driver compilation spikes, every target query runs an unrecorded warm-up pass. Reported data corresponds to the steady-state performance averaged over 10 subsequent kernel runs (variance under 10\%).

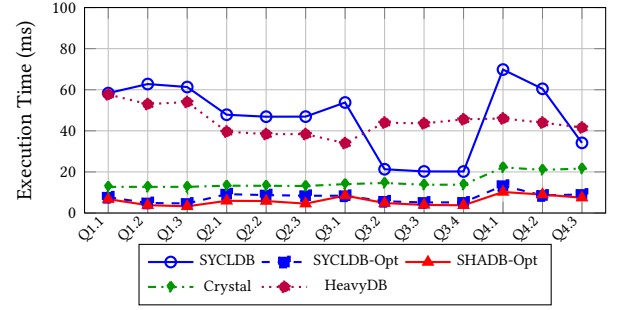
\begin{figure}[t!]
\centering
\begin{tikzpicture}
\begin{axis}[
    width=\columnwidth,
    height=4.3cm,
    ylabel={Execution Time (ms)},
    symbolic x coords={Q1.1, Q1.2, Q1.3, Q2.1, Q2.2, Q2.3, Q3.1, Q3.2, Q3.3, Q3.4, Q4.1, Q4.2, Q4.3},
    xtick=data,
    tick label style={font=\scriptsize},
    label style={font=\small},
    x tick label style={rotate=45, anchor=east},
    legend style={at={(0.5,-0.15)}, anchor=north, legend columns=3, font=\scriptsize},
    grid=major,
    ymin=0, ymax=100,
    enlarge x limits=0.05,
    title style={font=\small},
]
\addplot[color=blue, mark=o, thick]
coordinates {
(Q1.1,58.36) (Q1.2,62.77) (Q1.3,61.32)
(Q2.1,47.84) (Q2.2,46.88) (Q2.3,46.92)
(Q3.1,53.79) (Q3.2,21.31) (Q3.3,20.25) (Q3.4,20.21)
(Q4.1,69.82) (Q4.2,60.45) (Q4.3,34.14)
};
\addplot[color=blue, mark=square*, thick, dashed]
coordinates {
(Q1.1,7.69) (Q1.2,4.90) (Q1.3,4.65)
(Q2.1,9.18) (Q2.2,8.74) (Q2.3,8.42)
(Q3.1,8.41) (Q3.2,5.72) (Q3.3,5.15) (Q3.4,5.18)
(Q4.1,13.32) (Q4.2,8.55) (Q4.3,9.04)
};
\addplot[color=red, mark=triangle*, thick]
coordinates {
(Q1.1,6.76) (Q1.2,3.82) (Q1.3,3.27)
(Q2.1,5.94) (Q2.2,5.83) (Q2.3,4.57)
(Q3.1,8.49) (Q3.2,4.86) (Q3.3,3.94) (Q3.4,3.78)
(Q4.1,10.24) (Q4.2,8.98) (Q4.3,7.46)
};
\addplot[color=green!60!black, mark=diamond*, thick, dashdotted]
coordinates {
(Q1.1,12.74) (Q1.2,12.73) (Q1.3,12.74)
(Q2.1,13.30) (Q2.2,13.26) (Q2.3,13.16)
(Q3.1,14.07) (Q3.2,14.62) (Q3.3,13.78) (Q3.4,13.78)
(Q4.1,22.23) (Q4.2,21.01) (Q4.3,21.63)
};
\addplot[color=purple, mark=*, thick, dotted]
coordinates {
(Q1.1,57.70) (Q1.2,53.00) (Q1.3,54.00)
(Q2.1,39.60) (Q2.2,38.40) (Q2.3,38.40)
(Q3.1,34.00) (Q3.2,44.00) (Q3.3,43.60) (Q3.4,45.60)
(Q4.1,46.00) (Q4.2,44.00) (Q4.3,41.60)
};
\legend{SYCLDB, SYCLDB-Opt, SHADB-Opt, Crystal, HeavyDB}
\end{axis}
\end{tikzpicture}
\Description{A line graph plotting execution time in milliseconds for individual Star Schema Benchmark queries from Q1.1 to Q4.3 on an NVIDIA L40S, comparing the high execution times of baseline SYCLDB and HeavyDB against the significantly faster, low latency baselines of optimized SYCLDB-Opt, hand-written Crystal, and the lowest query-specific baseline, SHADB-Opt.}
\caption{SSB SF100 query execution time on \textbf{NVIDIA L40S} (kernel time,
warm run).}
\label{fig:end-to-end-three}
\end{figure}

\begin{figure}[t!]
\centering
\begin{tikzpicture}
\begin{axis}[
    width=\columnwidth,
    height=4.3cm,
    ylabel={Execution Time (ms)},
    symbolic x coords={Q1.1, Q1.2, Q1.3, Q2.1, Q2.2, Q2.3, Q3.1, Q3.2, Q3.3, Q3.4, Q4.1, Q4.2, Q4.3},
    xtick=data,
    tick label style={font=\scriptsize},
    label style={font=\small},
    x tick label style={rotate=45, anchor=east},
    legend style={at={(0.5,-0.15)}, anchor=north, legend columns=4, font=\scriptsize},
    grid=major,
    ymin=0, ymax=110,
    enlarge x limits=0.05,
    title style={font=\small},
]
\addplot[color=blue, mark=square*, thick, dashed]
coordinates {
(Q1.1,14.86) (Q1.2,9.31) (Q1.3,8.80)
(Q2.1,18.16) (Q2.2,16.17) (Q2.3,16.19)
(Q3.1,15.90) (Q3.2,10.48) (Q3.3,9.30) (Q3.4,9.35)
(Q4.1,26.55) (Q4.2,16.19) (Q4.3,16.93)
};
\addplot[color=red, mark=triangle*, thick]
coordinates {
(Q1.1,14.57) (Q1.2,7.45) (Q1.3,6.92)
(Q2.1,14.20) (Q2.2,13.33) (Q2.3,13.13)
(Q3.1,16.61) (Q3.2,8.33) (Q3.3,7.40) (Q3.4,7.34)
(Q4.1,17.08) (Q4.2,16.41) (Q4.3,13.39)
};
\addplot[color=green!60!black, mark=diamond*, thick, dashdotted]
coordinates {
(Q1.1,25.44) (Q1.2,25.45) (Q1.3,25.48)
(Q2.1,26.15) (Q2.2,26.08) (Q2.3,26.13)
(Q3.1,33.51) (Q3.2,29.59) (Q3.3,27.37) (Q3.4,27.40)
(Q4.1,83.22) (Q4.2,46.25) (Q4.3,43.57)
};
\addplot[color=purple, mark=*, thick, dotted]
coordinates {
(Q1.1,100.70) (Q1.2,97.30) (Q1.3,95.80)
(Q2.1,70.00) (Q2.2,67.00) (Q2.3,66.40)
(Q3.1,66.70) (Q3.2,69.00) (Q3.3,69.00) (Q3.4,68.40)
(Q4.1,94.00) (Q4.2,86.70) (Q4.3,84.00)
};
\legend{SYCLDB-Opt, SHADB-Opt, Crystal, HeavyDB}
\end{axis}
\end{tikzpicture}
\Description{A line graph plotting execution time in milliseconds for individual Star Schema Benchmark queries from Q1.1 to Q4.3 at SF200 scale on an NVIDIA L40S, comparing SYCLDB-Opt, the LLM-synthesized SHADB-Opt, hand-written Crystal, and HeavyDB. HeavyDB is the slowest across all queries, Crystal spikes sharply on Q4.3, while SYCLDB-Opt and SHADB-Opt remain consistently low.}
\caption{SSB SF200 query execution time on \textbf{NVIDIA L40S} (kernel time,
warm run).}
\label{fig:end-to-end-sf200}
\end{figure}
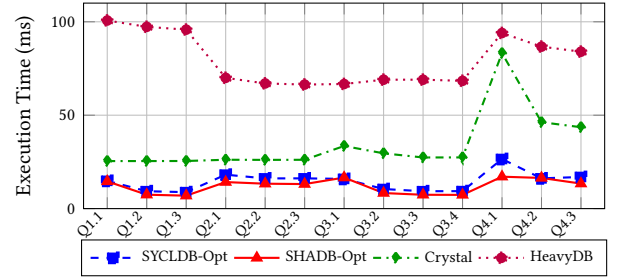

\subsection{Cross-System Comparison}

Figure~\ref{fig:end-to-end-three} shows the per-query execution time of fully-optimized SYCLDB together with baseline SYCLDB, SHADB, HeavyDB, and Crystal. There are several important observations to be made. 
First, SYCLDB-Opt totals $98.95$\,ms across the benchmark, achieving a $6.1\times$ improvement over the baseline ($604.06$\,ms). This demonstrates the effectiveness of fusion and the lifted optimizations. Second, SYCLDB-Opt outperforms both Crystal ($2.0\times$ faster) and HeavyDB ($5.9\times$ faster). Note here that unlike Crystal, which is a hard-coded implementation of SSB, SYCLDB is a general-purpose engine that can support star-schema workloads. This shows that portable design choices (flat instead of tiled, composition-based fusion instead of code-generation-based fusion) used by SYCLDB-Opt can provide competitive performance. Third, SYCLDB-Opt is only outperformed by SHADB ($77.94$\,ms total, $1.27\times$ faster). SYCLDB-Opt is within $1.4\times$ on nine of the thirteen queries. SYCLDB even matches synthesized code on Q3.1 ($8.41$ vs $8.49$\,ms) and Q4.2 ($8.55$ vs $8.98$\,ms). The gap widens to $1.8\times$ only on the part/supplier-join Q2.x cases (e.g.\ Q2.3, $8.42$ vs $4.57$\,ms). Baseline SYCLDB, in contrast, was $8\times$ slower than SHADB.  

Figure~\ref{fig:end-to-end-sf200} repeats this comparison at SF200 ($\sim$1.2B fact rows) to test whether these trends hold at larger scale. As can be seen, the relative ordering of the systems is similar at SF200. SYCLDB-Opt achieves near-linear scaling (total time of $188.19$\,ms, $1.9\times$ SF100) as the data doubles, and continues to outperform both Crystal ($445.64$\,ms, $2.4\times$ faster) and HeavyDB ($1035.0$\,ms, $5.5\times$ faster). Similar to SF100, SHADB outperforms SYCLDB-Opt only by a factor of $1.2\times$, confirming that the fusion and lifted optimizations remain effective and that the portable engine continues to track the specialized code across scale factors.

\paragraph{Compilation cost.}
Because SYCLDB composes and lowers its fused kernels at runtime, the first time it sees a new query shape it pays a one-time JIT cost. On the L40S, this JIT takes $0.40$\,s for a scalar scan and $0.84\text{--}1.05$\,s for a join. Profiling revealed that ${\sim}97\%$ of this time is the device-side lowering performed by AdaptiveCpp. Each compiled query shape is cached by AdaptiveCpp and the cost is never paid again for any query with a previously cached shape. Thus, steady-state execution time for these queries is just kernel time, which is the data that we report in this paper. Two observations place this one-time cost in perspective. First, runtime compilation is not unique to SYCLDB but intrinsic to any JIT-compiled database. HeavyDB, for instance, lowers each query fragment through an LLVM backend and pays an analogous one-time compile, which we likewise warm up and exclude from its reported times. Second, this is precisely the specialization cost that the synthesize-not-engineer approach pays in far heavier form. SYCLDB specializes a query shape in ${\sim}1$\,s  whereas SHADB's profile-guided synthesis of the same suite takes $4.2$\,GPU-hours and ${\sim}\$126$. Under the repetitive-query workloads that motivate query specialization in the first place, both are paid once and amortized, but the engineered one is roughly three orders of magnitude smaller.

\subsection{Cross-Architecture Performance Portability}

We now evaluate the performance-portability of SYCLDB by evaluating it on the AMD Instinct MI210 GPU. The only change we made to the configuration is to increase the number of items per thread from 4 to 16 to saturate the high bandwidth of HBM2e memory on the AMD GPU. Otherwise, we use the identical SYCLDB engine that is compiled with AdaptiveCpp to a single, portable binary (\texttt{--acpp-targets=generic}) that JIT-compiles the composition-based, fused kernels to the \texttt{amdgcn} target without any architectural modifications. We verified that the cross-architecture portability of the engine holds end-to-end by running the complete SSB SF100 suite with all optimizations enabled. Across all 13 queries in the SSB suite, the MI210 executes correctly and produces result fingerprints (MD5) that perfectly match the outputs from our reference runs, confirming the functional reliability of our portable fusion pipeline.


\textbf{Execution Performance.} To understand the impact of our generalizable optimizations on the AMD architecture, we compare the fully optimized SYCLDB engine on the MI210 against the SYCLDB baseline (\textit{SYCLDB base}). We also used SHADB to generate a native SHADB-Opt HIP ceiling for the AMD GPU. Across all 13 queries, the synthesized HIP code matched the correctness oracle. The profile-guided optimization loop successfully tuned the baseline kernels, yielding an improvement of up to $1.36\times$, with scan queries (Q1.x) primarily benefiting similar to the NVIDIA case reported in \S~\ref{sec:bg_how_good}. Profiling metrics from \texttt{rocprof} confirmed that the optimized scan kernels achieved a sustained HBM bandwidth of $0.95$\,TB/s ($58\%$ of the HBM2e max) by utilizing vectorized loads and memory-level parallelism. For join queries, the kernels saturated the vector memory unit at $95\%$--$98\%$ utilization, indicating they operate near the hardware limit for L2-resident random probes. These results demonstrate that the PGO loop generalizes effectively to ROCm/HIP, guiding the LLM to target device-specific bandwidth and memory execution floors. 

\begin{figure}[t!]
\centering
\begin{tikzpicture}
\begin{axis}[
    width=\columnwidth,
    height=4.3cm,
    ylabel={Execution Time (ms)},
    symbolic x coords={Q1.1, Q1.2, Q1.3, Q2.1, Q2.2, Q2.3, Q3.1, Q3.2, Q3.3, Q3.4, Q4.1, Q4.2, Q4.3},
    xtick=data,
    tick label style={font=\scriptsize},
    label style={font=\small},
    x tick label style={rotate=45, anchor=east},
    legend style={at={(0.5,-0.15)}, anchor=north, legend columns=3, font=\scriptsize},
    grid=major,
    ymin=0, ymax=60,
    enlarge x limits=0.05,
    title style={font=\small},
]
\addplot[color=blue, mark=o, thick]
coordinates {
(Q1.1,41.20) (Q1.2,46.40) (Q1.3,46.20)
(Q2.1,43.70) (Q2.2,40.50) (Q2.3,33.90)
(Q3.1,45.30) (Q3.2,38.60) (Q3.3,35.20) (Q3.4,35.00)
(Q4.1,57.70) (Q4.2,49.50) (Q4.3,48.00)
};
\addplot[color=blue, mark=square*, thick, dashed]
coordinates {
(Q1.1,6.96) (Q1.2,3.84) (Q1.3,3.46)
(Q2.1,19.21) (Q2.2,17.86) (Q2.3,17.56)
(Q3.1,16.71) (Q3.2,13.11) (Q3.3,11.11) (Q3.4,10.93)
(Q4.1,23.54) (Q4.2,19.03) (Q4.3,18.46)
};
\addplot[color=red, mark=triangle*, thick]
coordinates {
(Q1.1,6.01) (Q1.2,2.13) (Q1.3,1.73)
(Q2.1,16.77) (Q2.2,14.94) (Q2.3,14.69)
(Q3.1,16.99) (Q3.2,10.69) (Q3.3,8.82) (Q3.4,8.78)
(Q4.1,21.06) (Q4.2,17.27) (Q4.3,10.92)
};
\legend{SYCLDB, SYCLDB-Opt, SHADB-Opt}
\end{axis}
\end{tikzpicture}
\Description{A line graph plotting execution time in milliseconds for individual Star Schema Benchmark queries from Q1.1 to Q4.3 on an AMD MI210 GPU, demonstrating that the high execution times of the baseline SYCLDB engine are significantly reduced by the optimized SYCLDB-Opt variant, closely tracking the performance of the query-specific SHADB-Opt ceiling across all benchmarks.}
\caption{SSB SF100 query execution time on \textbf{AMD MI210} (kernel time,
warm run).}
\label{fig:end-to-end-amd}
\end{figure}
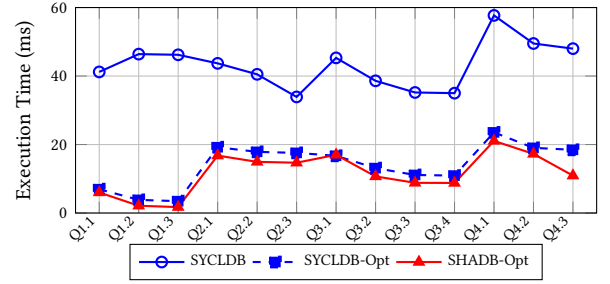

Figure~\ref{fig:end-to-end-amd} shows per-query execution time under SF100 on the AMD GPU. As can be seen, relative to the native SHADB-Opt HIP ceiling, SYCLDB-Opt is $1.2\times$ slower in total time ($181.78$\,ms versus $150.8$\,ms), while the unfused SYCLDB baseline is $3.7\times$ slower ($561.20$\,ms). In 10 out 13 queries, SYCLDB-Opt is less than $1.2\times$ slower than SHADB. Only in Q1.2, Q1.3, and Q4.3 is SYCLDB-Opt notably slower, by $1.8\times$, $2.0\times$ and $1.7\times$ respectively. These are the three queries whose kernels move the least data, and that drives the gap. Every fused kernel pays a fixed, per-row dispatch cost with dynamic functions. When a query does substantial memory work, like streaming many columns in a scan, or issuing many dependent probes in a join, this overhead is hidden behind that work, and SYCLDB-Opt stays close to SHADB. When little data is moved, the per-row dispatch instead dominates the runtime. SHADB does not pay this overhead, since it hard-codes everything. The three queries above trigger early filtering, where a highly-selective predicate eliminates many rows early in the operator chain. This results in little data being moved between operators exposing the dispatch overhead. 

Hardware profiling via \texttt{rocprof} (Table~\ref{tab:amd_profiling}) confirms this. In Q1.1, the scan fetches $4.78$\,GB at an achieved memory bandwidth of $0.85$\,TB/s, and runs at $96.6\%$ occupancy with near-zero stalls. The Q1.3 scan, in contrast, fetches only $2.29$\,GB as most rows fail to pass an early filter, leaving the dispatch overhead exposed. The joins split the same way. The low-selectivity Q4.1 keeps many survivors and is latency-bound on dependent random probes (memory stall $14.2\%$, $0.65$\,TB/s bandwidth usage). This is a hardware limit that SHADB's native HIP kernel hits as well. So SYCLDB-Opt stays within $1.1\times$. The high-selectivity Q4.3 drops nearly all rows and fetches just $5.0$\,GB. As a result, the memory unit is less utilized ($0.35$\,TB/s, $21\%$ of peak), and the dispatch overhead dominates, resulting in SYCLDB-Opt being $1.7\times$ behind SHADB. 

\begin{table*}[t]
\centering
\resizebox{0.85\textwidth}{!}{%
\begin{tabular}{llrrrrrr}
\toprule
\textbf{Query} & \textbf{Kernel} & \textbf{L2 Hit \%} & \textbf{Mem Stall \%} & \textbf{Mem BW (\% peak)} & \textbf{Fetched} & \textbf{VALU \%} & \textbf{Occupancy \%} \\
\midrule
Q1.1 (scan) & scalar terminal & 0.2 & 0.2 & 0.85 TB/s (52\%) & 4.78 GB & 26.2 & 96.6  \\
Q1.3 (scan) & scalar terminal & 0.8 & 0.8 & 1.13 TB/s (69\%) & 2.29 GB & 39.6 & 95.1  \\
Q4.1 (join) & group terminal & 81.4 & 14.2 & 0.65 TB/s (40\%) & 12.0 GB & 15.3 & 80.1  \\
Q4.3 (join) & group terminal & 91.0 & 0.5 & 0.35 TB/s (21\%) & 5.0 GB & 10.3 & 78.2  \\
\bottomrule
\end{tabular}
}
\caption{
Hardware counters for the dominant fused kernels on AMD MI210, opt-stack config (range fusion,
IPT=16). Mem BW is the achieved HBM read bandwidth (\% of the 1.6\,TB/s
peak in parentheses); Fetched is total HBM read volume.}
\label{tab:amd_profiling}
\end{table*}

\subsection{Optimization Microbenchmarks}

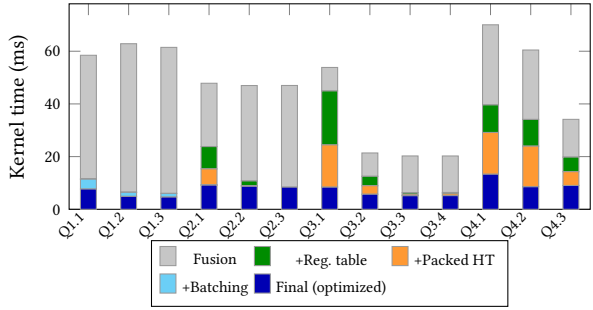
\begin{figure}[t]
\centering
\begin{tikzpicture}
\begin{axis}[
    width=\columnwidth,
    height=4.3cm,
    ybar stacked,
    bar width=6pt,
    ymin=0, ymax=78,
    ylabel={Kernel time (ms)},
    symbolic x coords={Q1.1,Q1.2,Q1.3,Q2.1,Q2.2,Q2.3,Q3.1,Q3.2,Q3.3,Q3.4,Q4.1,Q4.2,Q4.3},
    xtick=data,
    x tick label style={rotate=45, anchor=east, font=\scriptsize},
    tick label style={font=\scriptsize},
    label style={font=\small},
    legend style={at={(0.5,-0.15)}, anchor=north, legend columns=3, font=\scriptsize},
    reverse legend,
    enlarge x limits=0.04,
]
\addplot[fill=blue!70!black, draw=black!40] coordinates {
(Q1.1,7.69) (Q1.2,4.90) (Q1.3,4.65) (Q2.1,9.18) (Q2.2,8.74) (Q2.3,8.42)
(Q3.1,8.41) (Q3.2,5.72) (Q3.3,5.15) (Q3.4,5.18) (Q4.1,13.32) (Q4.2,8.55) (Q4.3,9.04)};
\addplot[fill=cyan!50, draw=black!40] coordinates {
(Q1.1,3.75) (Q1.2,1.54) (Q1.3,1.39) (Q2.1,0.08) (Q2.2,0) (Q2.3,0)
(Q3.1,0) (Q3.2,0) (Q3.3,0.08) (Q3.4,0) (Q4.1,0) (Q4.2,0) (Q4.3,0)};
\addplot[fill=orange!80, draw=black!40] coordinates {
(Q1.1,0.05) (Q1.2,0.02) (Q1.3,0) (Q2.1,6.08) (Q2.2,0.25) (Q2.3,0)
(Q3.1,16.03) (Q3.2,3.23) (Q3.3,0.52) (Q3.4,0.69) (Q4.1,15.84) (Q4.2,15.43) (Q4.3,5.24)};
\addplot[fill=green!55!black, draw=black!40] coordinates {
(Q1.1,0) (Q1.2,0) (Q1.3,0) (Q2.1,8.46) (Q2.2,1.76) (Q2.3,0)
(Q3.1,20.48) (Q3.2,3.57) (Q3.3,0.61) (Q3.4,0.51) (Q4.1,10.46) (Q4.2,10.13) (Q4.3,5.53)};
\addplot[fill=gray!45, draw=black!40] coordinates {
(Q1.1,46.97) (Q1.2,56.39) (Q1.3,55.42) (Q2.1,24.04) (Q2.2,36.23) (Q2.3,38.58)
(Q3.1,8.92) (Q3.2,8.86) (Q3.3,13.89) (Q3.4,13.86) (Q4.1,30.41) (Q4.2,26.35) (Q4.3,14.33)};
\legend{Final (optimized), {+}Batching, {+}Packed HT, {+}Reg.\ table, Fusion}
\end{axis}
\end{tikzpicture}
\Description{A stacked bar chart mapping the cumulative execution time reduction achieved by incrementally applying optimizations—Fusion, Register Table, Packed Hash Table, and Batching—to the baseline SYCLDB engine across individual Star Schema Benchmark queries on an NVIDIA L40S, with Fusion representing the largest performance gain across all workloads.}
\caption{Per-optimization decomposition of SYCLDB kernel time on the NVIDIA~L40S
(SF100).}
\label{fig:opt-progression}
\end{figure}

Having demonstrated the effectiveness and performance portability of all optimizations in SYCLDB, we now present an ablation study to isolate the relative improvement provided by each optimization. Figure~\ref{fig:opt-progression} shows the performance of variants of SYCLDB under SSB with optimizations added one at a time to the baseline version. 


\noindent Figure~\ref{fig:opt-progression} stacks this progression per query. The total execution time improves from the unfused $604.06$\,ms baseline to $98.95$\,ms ($6.1\times$), with each optimization acting on a distinct query class. \textbf{Fusion} is by far the largest
step ($604.06$ to $229.80$\,ms, $2.63\times$) and the only one that
helps every query. Collapsing the operator chain into a single kernel removes the
repeated streaming of the fact columns, making fusion the dominant (grey) band on
all 13 bars. The register table optimization improves join flights ($229.80$ to $168.56$\,ms,  $1.36\times$) by reducing memory access for intermediates. It moves the join-heavy Q3.1 ($44.87$ to $24.39$\,ms, $1.84\times$) and Q4.1 ($39.41$ to $28.95$\,ms) most, while the scalar Q1.x scans do not benefit from it. The packed hash table optimization is the largest of the three lifted optimizations ($168.56$ to $105.26$\,ms, $1.60\times$) and once again benefits joins predominantly (Q3.1 $24.39$ to $8.36$\,ms, Q4.2 $23.97$ to $8.54$\,ms). Finally, the per-terminal batching only improves the three scalar scan queries (Q1.x, ${\approx}1.4\times$ each), where streaming several rows per thread restores the bandwidth that a one-row-per-thread reduction left unused. 

\section{Related Work}

\paragraph{GPU database engines.}
Work on GPU-accelerated analytics has been explored across complete database engines, coprocessor architectures, and specialized query kernels. Systems such as HeavyDB~\cite{heavydb}, Ocelot~\cite{ocelot34}, CoGaDB~\cite{Bress2014}, Kinetica~\cite{kinetica40}, BlazingSQL~\cite{blazingsql12}, RAPIDS/cuDF~\cite{rapids67}, Theseus~\cite{theseus7}, together with other commercial and research systems~\cite{commercial64, others1, others15, others21, others31, others38, others68, others72} have shown that GPUs can accelerate relational operators but have also exposed practical bottlenecks such as data movement, kernel launch overhead, and operator-at-a-time materialization. These issues have been studied in both GPU-native databases and heterogeneous CPU/GPU execution environments 
~\cite{Cao2023, GPUanalytics, GPUmicrosftDB, Yuan2013, Li2016, Chrysogelos2019, Yogatama2022}. 

Recent systems have moved toward tighter integration between relational execution and accelerator runtimes. CoddSpeed~\cite{CoddSpeed} and the Tensor Query Processor line~\cite{TQP, tqp9, tqp14, tqp23, tqp35, tqp41, tqp79} map relational execution onto tensor and accelerator runtimes. BOSS~\cite{BOSS} and Sirius~\cite{sirius} compose GPU execution with existing engines and columnar interchange formats. Crystal~\cite{crystal} studies handwritten CUDA kernels as a performance reference for query-specialized execution. SHADB differs from these systems by using LLM synthesis to generate a complete query-specific CUDA/HIP implementation rather than relying on a fixed GPU operator library, a tensor runtime, or an LLVM/JIT backend.

\paragraph{LLM generation and synthesis.}
LLM-based code generation has recently been applied to kernels, systems, and database execution. CUDA-LLM~\cite{chen2025cudallm} and KernelBench~\cite{LLMkernelbench} show that frontier models can generate efficient CUDA kernels and benefit from execution/profiling feedback but also highlight correctness failures and nondeterminism. Jitskit~\cite{LLMjit} extends synthesis from isolated kernels to JIT-generated systems code, 
and DBPlanBench~\cite{LLMqueryplans} uses LLMs to modify physical query plans. In databases, GenDB~\cite{gendb} and BespokeOLAP~\cite{bespokeolap} synthesize workload-specific CPU engines, showing that LLMs can remove general-purpose database overhead when the target is a CPU execution model. SHADB focuses on the complementary GPU setting. It combines source audits, correctness validation, runtime diagnostics, and hardware profiling to prevent CPU fallbacks and guide LLM-generated kernels toward memory-bandwidth saturation. This places SHADB between hand-written GPU systems such as Crystal and CPU-oriented synthesis frameworks such as GenDB and BespokeOLAP.

Existing work has explored synthesized query execution on CPUs, hand-engineered GPU execution engines, and performance-portable database systems as largely independent directions. In contrast, this work investigates whether LLM-synthesized GPU implementations can serve as a vehicle for discovering general optimizations that transfer back into a reusable, performance-portable query engine.

\section{Discussion \& Conclusion}

In this paper, we first looked at how effective LLMs are in generating GPU code. In order to do so, we presented SHADB, a framework for synthesizing query-specific GPU database kernels using LLMs and profile-guided feedback. Using SHADB, we showed that LLMs, when guided by a PGO loop, can generate customized GPU code that outperforms engineered CPU and GPU baselines by effectively exploiting the GPU memory bandwidth. Second, we performed an in-depth analysis of SHADB-synthesized code to identify optimizations that gave it an edge over engineered systems and categorized these optimizations into generalizable ones that can be lifted into a database engine without compromising portability, and workload/hardware-specific ones that are specific to the given setting and non-portable. Third, we took SYCLDB, a standards-based, performance-portable, prototype database engine and systematically added optimizations to overcome various bottlenecks inherent to the operator-at-a-time execution model of the baseline. In doing so, we showed how dynamic functions provided by AdaptiveCpp compiler enables composition-based fusion as an alternative to code-generation-based fusion. Finally, by evaluating optimized SYCLDB, we showed how it is possible to engineer optimizations that bridge the gap with synthesized code without sacrificing generality or portability.

Taken together, our results lead us to conclude that, on GPUs, LLM-based synthesis is unlikely to displace carefully engineered query engines. The marginal benefit of synthesized code is small; across the SSB suite at SF100, SHADB is only $1.27\times$ faster than optimized SYCLDB. This performance improvement is achieved at a steep and recurring cost, as synthesizing and profile-tuning the suite consumes roughly \$126 of model calls and several GPU-hours. The resulting code is also fundamentally non-portable, as the very specializations that yield the speedup (encoding dataset cardinalities and key ranges as compile-time constants, hand-fitting kernels to one device, etcetera) are also the ones that destroy generality. Engineered code carries no such liability. As we showed, the same SYCLDB engine executes the full suite on both an NVIDIA L40S and an AMD MI210. It is also designed to support unseen star-schema queries.

Our statement about engineering being the preferred choice stands in direct contrast to recent CPU-targeted work. On CPUs, synthesized systems such as GenDB and Bespoke OLAP outperform general engines like DuckDB by more than an order of magnitude, because a CPU engine's interpretation overhead may be large enough to justify discarding generality for query-specific code. On GPUs, the dominant cost is instead memory bandwidth, and once a portable engine saturates it through fusion and the lifted optimizations, the residual gap to specialized code is too small to justify its dollar cost and its lack of generality and portability. 

We do, however, see a constructive role for LLMs as a discovery tool for optimizations. Similar to recent work on optimizing CPU-based engines using LLM-extracted primitives~\cite{tailwind}, our work lifted two optimizations into SYCLDB that were identified by SHADB's profile-guided loop. More such workload-specific optimizations could in principle be lifted into a general engine as specialized code paths activated for particular query shapes or hardware. However, in our experience, this comes at high software-engineering cost, as every specialized path is an additional branch that must be kept correct with proper fallback, potentially tuned across scale factors, and query variants, and periodically checked to avoid dead code. Whether the speedup from a specialized path justifies the ongoing burden of maintaining yet another divergent code path alongside the portable default is a tradeoff that future database developers have to reconcile.



\bibliographystyle{ACM-Reference-Format}
\bibliography{sample}

\end{document}